\documentclass[aps,amsfonts,nofootinbib,twocolumn]{revtex4-1}
\usepackage{amsmath}
\usepackage{amstext}
\usepackage{amsopn}
\usepackage{amsfonts}
\usepackage{amssymb}
\usepackage{bbm}
\usepackage{accents}
\usepackage{empheq}
\usepackage{graphicx}
\usepackage{epsf}
\usepackage{epsfig}
\usepackage{graphics}
\usepackage[latin1]{inputenc}
\newcommand{\be}{\begin{equation}}
\newcommand{\ee}{\end{equation}}
\newcommand{\bea}{\begin{eqnarray}}
\newcommand{\eea}{\end{eqnarray}}
\def\k{\mathbf{k}}
\def\x{\mathbf{x}}
\def\s{\mathbf{s}}

\def\p{\mathbf{p}}
\def\r{\mathbf{r}}
\def\S{\boldsymbol \sigma}
\begin{document}
\title{Spin Triplet Pairing for Superconductivity}

\author{Ashok Das}
\affiliation{Department  of  Physics   and  Astronomy,  University  of
Rochester, Rochester, NY 14627-0171, USA}
\affiliation{Saha Institute of Nuclear Physics, 1/AF Bidhannagar, Calcutta 700064, India}
\author{J. Gamboa}
\affiliation{Departamento de F\'{\i}sica, Universidad de Santiago de Chile, Casilla 307, Santiago, Chile}
\author{ F. M\'endez}
\affiliation{Departamento de F\'{\i}sica, Universidad de Santiago de Chile, Casilla 307, Santiago, Chile}
\author{F. Torres}
\affiliation{Departamento de F\'{\i}sica, Universidad de Santiago de Chile, Casilla 307, Santiago, Chile}

\begin{abstract}
A generalization of the Cooper pairing mechanism is proposed which allows for a triplet state of lower energy. This is achieved by incorporating spin into the canonical commutation relations and by modifying the $\delta$ potential contact interaction. The gap equation contain as solutions both singlet and triplet states. It is shown that the triplet state is lower in energy than the singlet state which may explain the spin-triplet superconductivity observed in heavy fermion compound UPt3 and in Sr$_2$RuO$_4$. \vskip 1.0cm \centerline{\it Dedicated to the memory of Olivier Espinosa}
\bigskip
\end{abstract}


\maketitle

Superconductivity \cite{Cooper,BCS} is a fascinating subject which has been studied vigorously both theoretically as well as experimentally in the past several decades \cite{degennes,tinkham}. In  conventional superconductivity, the electron-phonon interaction is known to lead to a condensed state of $s$-wave Cooper pairs and this has been experimentally verified. However, in certain materials such as the heavy fermion compound UPt3 \cite{expe1,expe2} as well as Sr$_2$RuO$_4$ \cite{rice,expe3}, it has been observed that the Cooper pairs occur in a triplet state. This unconventional superconductivity \cite{benne,pita} would be compatible with some  kind of ferromagnetic effect due to the spin in a manner analogous to the spin triplet state in superfluid He$^3$. However, there is no obvious way to accommodate this within the framework of the conventional description of superconductivity. Any generalization of the Cooper pair mechanism should not only allow for the existence of both singlet as well as triplet states, but should also predict that the triplet state is energetically favorable at low energies.
 
In the absence of a mechanism to describe the unconventional superconductivity within the usual framework of  quantum mechanics, in this letter we propose a minimal modification of quantum mechanics that can achieve this which can be thought of as a starting point for a macroscopic theory for such systems. Basically we modify the canonical commutators in such a way that the space coordinates and spin become coupled through 
 \be 
 [\hat{x}_i,\hat{x}_j]= - i \theta^2\epsilon_{ijk}s_k,\ [\hat{x}_{i}, s_{j}] = -\theta \epsilon_{ijk} s_{k}, \label{1}
 \ee 
while all the other commutators involving spin and momentum remain unchanged. Here $\theta$ is a very small real constant parameter with dimensions of length. In fact, such a modified algebra was already proposed in  \cite{FGLMP} and studied for the case where the parameter $\theta$ was real. Equation \eqref{1} merely corresponds to the case where this parameter is purely imaginary (namely, letting $\theta\rightarrow i\theta$ in \cite{FGLMP}) and in this case the operator $\hat{x}_{i}$ ceases to be Hermitian. As a result, the full algebra satisfying Jacobi identity also involves the additional relations  
\begin{align}
\label{algdag}
& [{\hat{x}_i}^\dag,{\hat{x}_j}^\dag] = -i \theta^2\epsilon_{ijk}s_k,&  &[{\hat{x}_i}^\dag,\hat{p}_j]=i\delta_ {ij}, \nonumber\\ 
& [{\hat{x}_i}^\dag,s_j] = \theta \epsilon_{ijk}s_k,&  &[{\hat{x}_i}^\dag,\hat{x}_j]= i \theta^2\epsilon_{ijk}s_k.
 \end{align}  

It is clear from \eqref{1} and \eqref{algdag} that the coordinate operators $\hat{x}_{i}$ are now matrices in the spin space. Furthermore, the complete algebra can also be realized in terms of the conventional operators through the change of basis
\be 
{\hat x}_i = x_i  + i \theta s_i,\quad {\hat{x}_{i}}^{\dagger} = x_{i} - i \theta s_{i},\quad \hat{p}_i = p_i. \label{bopp}
\ee
where $(x_{i},p_{i},s_{i})$ satisfy the standard commutation relations. In order to generalize the Cooper pair mechanism, we note that given a real (Hermitian) potential $V (\mathbf{x})$, a naive generalization to a standard Hamiltonian of the form (we set $m=1$)
\be
H (\hat{\mathbf{x}},\hat{\mathbf{p}}) = \frac{\hat{\mathbf{p}}^{2}}{2} + V (\hat{\mathbf{x}}),
\ee
would no longer be Hermitian nor would it be ${\cal PT}$ symmetric \cite{PT} where the transformations of $\hat{x}_{i}$ under ${\cal P}$ and ${\cal T}$ can be seen from \eqref{bopp} to correspond to
\be
\hat{x}_{i}\xrightarrow{\cal P} -{\hat{x}_{i}}^\dag,\  
{\hat{x}_i}^\dag\xrightarrow{\cal P} -\hat{x}_i;\  \hat{x}_{i}\xrightarrow{\cal T} {\hat{x}_{i}},\  
{\hat{x}_i}^\dag\xrightarrow{\cal T} {\hat{x}_i}^\dag. \label{PT}
\ee
We note here that even a ${\cal PT}$ symmetric potential $V (i\mathbf{x})$ would not generalize to ${\cal PT}$ symmetric system and we would restrict ourselves to real potential $V (\mathbf{x})$ in this work since our goal is to generalize the Cooper pair mechanism.  In order to ensure the reality of the spectrum of the Hamiltonian, we note that given a potential $V (\hat{\mathbf{x}})$, we can construct two Hermitian Hamiltonians in a straightforward manner as \cite{bazeia}
\bea 
\label{HNC+}
H_{+}(\hat{\x},\hat{\p})&=& \frac{\hat{\p}^{2}}{2} +
 \frac{1}{2}\left(V(\hat{\x}) + V^{\dagger}(\hat{\x}) \right) ,\\
\label{HNC-}
H_{-}(\hat{\x},\hat{\p})&=& \frac{\hat{\p}^{2}}{2} +
 \frac{i}{2} \left(V (\hat{\x})- V^{\dagger}(\hat{\x})\right).
\eea 
Furthermore, we note that  if  $V(-\hat{\x})=V(\hat{\x})$ then 
$H_{+}$ is also ${\cal PT}$ symmetric while if 
$V(-\hat{\x})=-V(\hat{\x})$ then $H_{-}$
is ${\cal PT}$  symmetric as well.  A straightforward 
calculation show that the  Hamiltonians (\ref{HNC+}) and (\ref{HNC-}) can be expressed in terms of the standard operators \eqref{bopp} as  
\begin{align}
H_{+}(\x,\p,\s) & = \frac{\p^{2}}{2} + \frac{1}{2}\left(V (\x+i\theta\s)
 + V(\x- i\theta \s)\right)\notag\\
 & = \frac{\p^{2}}{2} + \cos(\theta \s\cdot\mbox{\boldmath$\nabla$})\,V(\x),\label{H+}\\
H_{-}(\x,\p,\s) & = \frac{\p^{2}}{2} + \frac{i}{2}\left(V (\x+i\theta\s) - V(\x-i\theta \s)\right)\notag\\
& = \frac{\p^{2}}{2} - \sin(\theta\s\cdot\mbox{\boldmath$\nabla$})\,V(\x),\label{H-}
\end{align} 
where we have used the reality of the potential $V (\x)$. We note from \eqref{H+} and \eqref{H-} that when $\theta \rightarrow 0$
\be
H_{+}\rightarrow \frac{\p^{2}}{2}+ V(\x),\quad H_{-}\rightarrow \frac{\p^{2}}{2}.
\ee 
Therefore, we would concentrate on the Hamiltonian $H_{+}$ which would reduce in the limit $\theta\rightarrow 0$ to the conventional interacting Hamiltonian of the system. Furthermore, in spite of the nontrivial mixing of the space and spin variable, $H_{+}$ in \eqref{H+} commutes with $\s$ so that the eigenstates can be written as product states. 

The generalization of this construction to a many particle system can be
carried out in a straightforward manner.  For example, let us  consider a two particle system (which is relevant for our analysis) where the interaction depends on the relative coordinate $\r=\x_{1}-\x_{2}$. First, we note that the only combination of the coordinates $\hat{\x}_1$ and $\hat{\x}_2$ that commutes with the total spin $\mathbf{S}=\s_1+\s_2$ of the system is $\hat{\x}_1 - {\hat{\x}_2}^\dag$ and its conjugate. Therefore, the Hermitian potential which would also commute with the total spin can be written in the form
\begin{align}
V_{2} (\hat{\x}_{1},\hat{\x}_{2}) & = \frac{1}{2}\left( V(\hat{\x}_{1}-
\hat{\x}^{\dagger}_{2})+V^{\dagger}(\hat{\x}_{1}-
\hat{\x}^{\dagger}_{2})\right)\notag\\
& = \frac{1}{2}\left(V (\r+i\theta\mathbf{S}) + V (\r - i\theta\mathbf{S})\right),
\end{align}
where we are assuming that the couplings are real. In particular, if we consider  a  contact
potential which is responsible  for the  instability of  Fermi sea
against small attractive interactions between electrons
\cite{Cooper, BCS,pita}, this would generalize in our case to a  problem  of two  electrons
interacting through the \lq\lq contact"  potential 
\be
V_{2}(\r,\mathbf{S})  = -\frac{\gamma}{2}\left( \delta_{\theta} (\r + i\theta\mathbf{S})+
\delta_{\theta} (\r -i\theta \mathbf{S})\right)
\ee
where we have defined a modified delta function as
\be
\delta_{\theta} (\mathbf{z}) = \int\limits_{|\mathbf{k}|=0}^{|\mathbf{k}|=\frac{1}{\theta}} \frac{d^{3}k}{(2\pi)^{3}}\, e^{i\mathbf{k}\cdot \mathbf{z}},
\ee
which reduces to the standard delta function in the limit $\theta\rightarrow 0$. As a result, we can write the \lq\lq contact" potential as
\be
V_{2}(\r, \mathbf{S}) =  -\gamma \int\limits_{|\mathbf{k}|=0}^{|\mathbf{k}|=\frac{1}{\theta}} \frac{d^{3}k}{(2\pi)^{3}} \,e^{i\k\cdot\r} \cosh(\theta\k\cdot\mathbf{S}),\label{NCdelta}
\ee
where $\gamma$ is a positive constant. We note that the minimal modification of the delta potential 
preserves the qualitative nature of the electron-phonon interaction (namely, the two electrons interact through a local deformation of the lattice), but introduces a finite range of interaction of the order $2\theta$.

Since the potential depends on the relative coordinate, the energy levels are determined from the Schr\"{o}dinger equation for the reduced system
\be
\label{NCSch}
(- \mbox{\boldmath$\nabla$}^{2} + V_{2}(\r,\mathbf{S}))\Psi^{(s,m)} (\r) = E^{(s,m)}\Psi^{(s,m)} (\r),
\ee
where we use the compact notation $\Psi^{(s,m)} (\r) = \Psi (\r) |s,m\rangle$ with $|s,m\rangle$ labeling the representation of $\mathbf{S}$. In the momentum space
\be
\Psi^{(s,m)}(\r) = \int d^{3}k\, e^{i\k\cdot\r}\Phi^{(s,m)}(\k),\label{FT}
\ee
we can write \eqref{NCSch} as (for $|\mathbf{k}-\mathbf{k}'| < \frac{1}{\theta}$) 
\begin{widetext}
\begin{align}
\Phi^{(s,m)}(\k) & = \frac{\gamma}{(\k^{2}-E^{(s,m)})}\int d^{3}k' 
\cosh(\theta(\k-\k')\cdot\mathbf{S})\Phi^{(s,m)}(\k')\notag\\
& = \frac{\gamma}{(\k^{2}-E^{(s,m)})}\int d^{3}k' 
\left(\cosh(\theta\k\cdot\mathbf{S})\cosh(\theta\k'\cdot\mathbf{S})
+\sinh(\theta\k\cdot\mathbf{S})\sinh(\theta\k'\cdot\mathbf{S}) \right) \Phi^{(s,m)}(\k').\label{EqforE}
\end{align}
\end{widetext}
From \eqref{FT} we note that the antisymmetry of the two electron wavefunction 
under the exchange of particles determines ($P_{12}$ is the exchange operator) 
\begin{align}
P_{12} \Phi^{(s,m)} (\k) & = (-1)^{s+1} \Phi^{(s,m)} (\k),\notag\\
\Phi^{(s,m)} (-\k) & = (-1)^{s} \Phi^{(s,m)} (\k).
\end{align}

Using these relations, it follows from \eqref{EqforE} for the singlet state $(s=0)$ that  
\be
\label{Esin}
\gamma\int \frac{d^{3}k}{(2\pi)^{3}}\, \frac{\langle\cosh^{2}(\theta\k\cdot\mathbf{S})\rangle^{(0,0)}}{(\k^{2}-E^{(0,0)})}=1,
\ee
while the triplet state $(s=1)$ leads to  
\be
\label{Etri}
\gamma\int \frac{d^{3}k}{(2\pi)^{3}}\, \frac{\langle\cosh^{2}(\theta\k\cdot\mathbf{S})-1\rangle^{(1,m)}}{(\k^{2}-E^{(1,m)})}=1,
\ee
which can be thought of as the gap equations for the singlet and the triplet states respectively. Here $\langle \cdot \rangle^{(s,m)}$ represents the expectation value in the state $|s,m\rangle$. We can simplify these equations by noting that  
\be
(\k\cdot\mathbf{S})^{2} =k^{2}\left(1+(\k\cdot\S_{1})(\k\cdot\S_{2})k^{-2}\right)\!/2 = k^{2}S_{12},
\ee
where we have identified $k = |\k|, \s_{i} = \S_{i}/2, i=1,2$ and we note that $S_{12}^{2} = S_{12}$ which leads to 
\be
(\k\cdot\mathbf{S})^{2n}=k^{2n}S_{12},\ \cosh^{2}(\theta\k\cdot\mathbf{S})=1 + \sinh^{2}(\theta k)S_{12}.\label{simplification}
\ee
With a little bit of algebra, we can show that
\begin{align}
\int \frac{d\Omega}{4\pi}\,\langle S_{12}\rangle^{(0,0)} & = 0,\notag\\
\int \frac{d\Omega}{4\pi}\,\langle S_{12}\rangle^{(1,m)} & = \rho^{(m)} = \frac{1}{2} (1 + \frac{1}{6} (1- (-1)^{|m|})).\label{rho}
\end{align}
It is worth noting here that in the triplet state the expectation value in the state $m=0$ is lower than that in the states $m=\pm 1$ state which are degenerate and reflect a sort of ferromagnetic degeneracy.

To investigate the Cooper pair mechanism, we convert the momentum integrations in \eqref{Esin} and \eqref{Etri} to energy integrals through the identification $2\epsilon = k^{2}$. Furthermore, as is well known the electrons which contribute to superconductivity lie in a thin energy shell near the Fermi surface, namely, $\epsilon_{F} < \epsilon < \epsilon_{F} + \epsilon_{c}$ where $\epsilon_{F}$ is the Fermi energy and  $\epsilon_{c} \ll \epsilon_{F}$ so that we can restrict the energy integration to this small interval. Using \eqref{simplification} and \eqref{rho} in  (\ref{Esin}), we obtain
\be
\gamma N (0) \int\limits_{\epsilon_{F}}^{\epsilon_{F}+\epsilon_{c}}  \frac{d\epsilon}{(2\epsilon - E^{(0,0)})}=1,
\ee
where $N(0)=\frac{k^{2}}{2\pi^{2}} \frac{dk}{d\epsilon}$ is the density of states per unit energy interval which is assumed to be constant within this small range of the integral.
This (gap) equation for the singlet state coincides with the standard commutative case. Therefore,  there is no noncommutative correction to the energy of the singlet state which has the value
\be
E^{(0,0)} = 2\left(\epsilon_{F} - \epsilon_{c} e^{-\frac{2}{\gamma N (0)}}\right).\label{standard}
\ee

On the other hand, using \eqref{simplification} and \eqref{rho} in \eqref{Etri} we obtain 
\begin{align}
& \gamma \rho^{(m)}N(0)\int\limits_{\epsilon_{F}}^{\epsilon_{F}+\epsilon_{c}} d\epsilon\, \frac{\sinh^{2}(\theta \sqrt{2\epsilon})}{(2\epsilon- E^{(1,m)})}\notag\\
&\quad  \simeq 2\gamma \rho^{(m)}N(0) \theta^{2} \int\limits_{\epsilon_{F}}^{\epsilon_{F}+\epsilon_{c}}\frac{d\epsilon\, \epsilon}{(2\epsilon - E^{(1,m)})} =1.
\label{Etri2}
\end{align}
The integration can be carried out in a straightforward manner and leads to 
\be
\frac{2}{E^{(1,m)}} \left( \frac{1}{\gamma N(0) \rho^{(m)}\theta^{2}}-\epsilon_{c}\right) = 
\ln \left( 1+\dfrac{2\epsilon_{c}}{2\epsilon_{F}- E^{(1,m)}}\right). \label{trans}
\ee

To show that the triplet state has a lower energy than the singlet state, let us define the dimensionless ratio
\be
x = \frac{E^{(1,m)}}{E^{(0,0)}},\label{ratio}
\ee
and rewrite \eqref{trans} as 
\be
C = 1 + B x\,\ln\left(1 + \frac{1}{B(1-x) + e^{-A}}\right),\label{dimless}
\ee
where we have defined three dimensionless constants
\be
A = \frac{2}{\gamma N(0)},\ B = \frac{E^{(0,0)}}{\epsilon_{c}} = \frac{\epsilon_{F}}{\epsilon_{c}} -e^{-A},\ C=\frac{A}{2\rho^{(m)} \theta^{2}\epsilon_{c}}.
\ee
Equation \eqref{dimless} can be solved graphically by looking for the simultaneous solutions of the equations
\be
y = C,\ y = 1 + B x \ln\left(1 + \frac{1}{B(1-x) + e^{-A}}\right),\label{dimless1}
\ee

\begin{figure}[ht!]
\begin{center}
\includegraphics[scale=0.9]{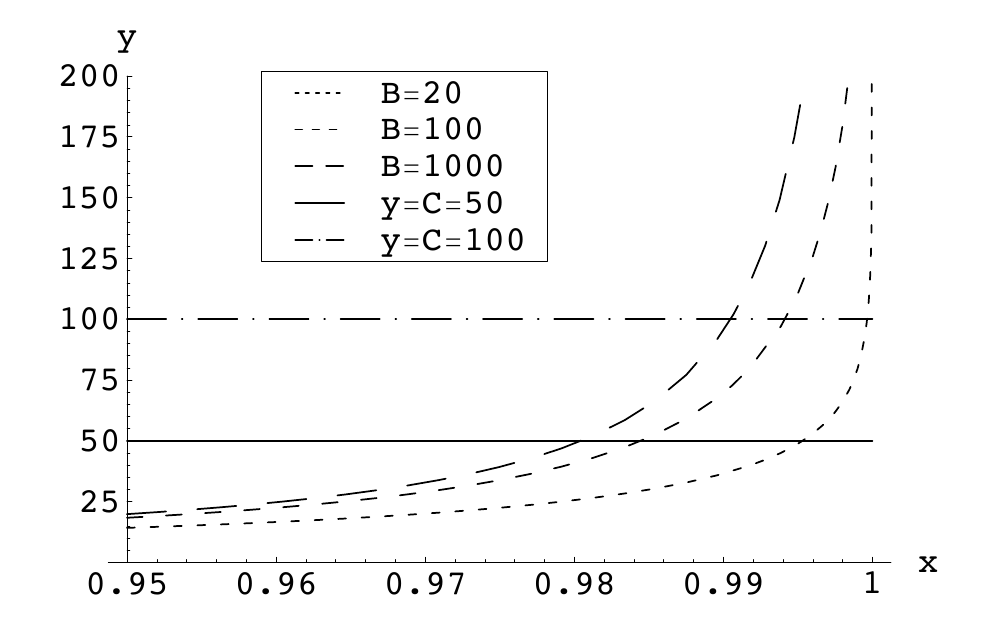}
\end{center}
\caption{Simultaneous solution of \eqref{dimless1}.}
\label{1}
\end{figure}
\noindent which are plotted in Fig. {1} for different values of the constants $B,C$. We note that in the weak coupling approximation $A \gg 6.7$ and for conventional superconductors $B > 10$ (recall that $\epsilon_{c}\ll \epsilon_{F}$) and since $\theta$ is a small parameter, the constant $C$ can be chosen to be of the order of $10^{2}$. The parameters in the plot are chosen keeping this in mind. Let us discuss some of the general features of the simultaneous solution. First as $\theta$ is taken to be smaller and smaller and, therefore, $C$ larger, the solution $x\rightarrow 1$ implying that in the limit $\theta\rightarrow 0$, the singlet and the triplet solutions become degenerate. However, for any finite and small $\theta$, the solution exists for $x < 1$ implying that the triplet solution is lower in energy than the singlet solution. Furthermore, from the fact that $\rho^{(m)}$ is larger for $m=\pm 1$ than for $m=0$ (see \eqref{rho}), the constant $C$ correspondingly is smaller for $m=\pm 1$ than for $m=0$. This implies that the triplet state with $m=\pm 1$ has the lowest (degenerate) energy. The graphical solutions can also be understood qualitatively as follows. Since $e^{-A}$ is a very small quantity, when $e^{-A} \ll B(1-x) < 1$, then \eqref{dimless} can be written approximately as
\be
C \approx \frac{1}{1-x},\quad {\rm or}\quad  x \approx 1 - \frac{1}{C} < 1.\label{analytic}
\ee
For $\theta\rightarrow 0, C\rightarrow \infty$ so that $x\rightarrow 1$ as we have already mentioned. For $m=\pm 1$, the value of $C$ is smaller than that for $m=0$ and, correspondingly, the value of $x$ would be lower for $m=\pm 1$.

The result of this analysis is quite interesting because it describes a theoretical mechanism for a triplet pairing state which is lower in energy than the singlet state which may explain the behavior of some of the unconventional superconductors. Although the precise mechanism of why some materials prefer a singlet state rather than a triplet state as ground state or vice versa is not understood yet, our analysis suggests that such an understanding necessarily requires us to  go beyond the conventional quantum theory.

\bigskip

\noindent{\bf Acknowledgment}
Two of the authors (JG and FT) would like to thank the kind hospitality at the University of Rochester where part of this work was done. This work was supported in part by US DOE Grant number DE-FG 02-91ER40685, by  MECESUP FSM 0605 and by FONDECYT-Chile grant-1095106, 1100777 and Dicyt (USACH).

\end{document}